\documentclass[prb,twocolumn,showpacs,floats,eqsecnum,amsmath,amssymb]{revtex4}
\usepackage[dvips]{graphicx}
\begin{document}

%\draft \twocolumn[\hsize\textwidth
%\columnwidth\hsize\csname
%@twocolumnfalse\endcsname

\title {Lower bound for ground state energy of BEC in a rotating optical lattice}

\author {Y. Azizi}

\affiliation{ Institute for Advanced Studies in Basic Sciences,
Zanjan 45195-1159, Iran}
\begin{abstract}
We use the frustrated $XY$ model approximation of BEC in a
rotating optical lattice and formulate the problem of the ground
state in terms of eigenvectors and eigenvalues of frustrated
adjacency matrix (coupling matrix). By using this formulation, we
show that there is a lower bound for ground state energy in terms
of maximum eigenvalue of this matrix.
\end{abstract}

\maketitle
\section{Introduction}
Effect of the rotation of optical lattice on physical properties
of BEC is an interesting problem. This effect can be observed as
vortex in lattice and in this way is related to many problems in
condensed matter physics, e.g. Josephson junctions arrays, ... .
\cite{R1}

It can be shown that with appropriate approximations
\cite{R2,R3,R4}, the problem of the ground state of BEC in a
rotating optical lattice can be formulated as frustrated $XY$
model. In this formulation, rotation plays the role of frustration
parameter\cite{R2,R3,R4}.

Here, we focus on ground state of BEC in a rotating optical
lattice in this approximation and by use of properties of
frustrated $XY$ model, we find a lower bound for ground state
energy for different optical lattices.

\section{BEC in rotating optical lattice: frustrated $XY$ model}

It can be shown that in appropriate approximation, Hamiltoian for
BEC in a rotating optical lattice can be reduce to frustrated $XY$
Hamiltonian\cite{R2,R3,R4},

\begin{eqnarray}
H=-\sum_{<i,j>}J_{ij}cos(\theta_i-\theta_j+A_{ij})
\end{eqnarray}
Suppose that $n_i$ is number of atoms in $i$th site and
$\theta_i$ is the phase of localized wave function on $i$th site.
$J_{ij}$ is proportional to $\sqrt{n_in_j}$ and $A_{ij}$ is line
integral of $A=\Omega\hat{z}\times r$. If we define frustration
parameter $f$ as $f=2\Omega m/h$, where $m$ is mass of condensate
and $h$ is plank constant, then we can define $2\pi
B_{ij}f=A_{ij}$ for nonzero $f$ and
\begin{eqnarray}\label{ham1}
H=-\sum_{<i,j>}J_{ij}cos(\theta_i-\theta_j+2\pi B_{ij}f)
\end{eqnarray}
Hamiltonian \ref{ham1} can be written as
\begin{eqnarray}\label{ham2}
H=-Re\{\sum_{<i,j>}J_{ij}exp(i(\theta_i-\theta_j+2\pi B_{ij}f))\}
\end{eqnarray}
where $Re\{x\}$ is real part. By introducing frustrated adjacency
matrix, $S_{ij}=J_{ij}exp(i2\pi B_{ij}f)$, we have,
\begin{eqnarray}\label{ham3}
H=-N/2Re\{\vec{\xi^*}S\vec{\xi}\}
\end{eqnarray}
where $N$ is number of sites and
$\xi_i=exp(-i\theta_i)/\sqrt{N}$. The factor $2$ enters because
we consider both $(i,j)$ and $(j,i)$ terms in this case.

Now we focus on formulating of the problem in terms of eigenvalues
and eigenvectors of $S$. Suppose that $\eta^k$ is eigenvector
correspond to eigenvalue $\lambda^k$ of matrix $S$, i.e.
$S\eta^k=\lambda^k\eta^k$. Also we suppose that set of
eigenvectors of $S$ are complete and orthogonal. Then we can write
following expansion for $\xi$,
\begin{eqnarray}\label{exp1}
\xi=\sum_{k}c^k\eta^k
\end{eqnarray}
Using this expansion in Hamiltonian \ref{ham3},
\begin{eqnarray}\label{ham4}
H=-N/2Re\{\sum_kc^kc^{k*}\lambda_k\}
\end{eqnarray}
But there is some restriction on $c$ coming from properties of
$\xi$. Because of $\xi.\xi^*=1$, we have $\sum_kc^kc^{k*}=1$ and
therefore $|c^k|\le 1$ for any $k$. Also from definition of
$\xi_l$ we have,
\begin{eqnarray}
\xi_l\xi_l^*=\frac{1}{N}
\end{eqnarray}
which leads to,
\begin{eqnarray}\label{res1}
\sum_{m,k}c^mc^{k*}\eta_l^m\eta_l^{k*}=\frac{1}{N}
\end{eqnarray}

The problem of minimization of Hamiltonian \ref{ham3} is
equivalent to minimization of Hamiltonian \ref{ham4} with respect
to complex numbers $c_i$'s and with constrains \ref{res1}.

\section{Lower bound for ground state energy}
We focus on minimization problem defined by equations \ref{ham4}
and restrictions \ref{res1}.

We consider simple situation, a plaquette with unit length and
uniform $J_{ij}$, which can be shown that frustrated $XY$ model
for it can be solved exactly. In this case, $S$ is given as,
$S_{12}=1,S_{23}=1,S_{34}=exp(-2\pi i f),S_{41}=1$, and
$S_{ji}=S^*_{ij}$ (which is property of $S$). Numerical results
shows that in this case, $E/N=-\lambda_{max}/2$.

From the general Hamiltonian \ref{ham4} and the constrain's on
$c^i$'s ($|c^i|\le 1$), we can see that,

\begin{eqnarray}
H/N\ge -\lambda_{max}/2
\end{eqnarray}
Therefore the minimum of Hamiltonian also satisfies this
inequality. It means that
\begin{eqnarray}
E/N\ge -\lambda_{max}/2
\end{eqnarray}

\section{conclusion}
We have shown that ground state energy of a BEC in an rotating
optical lattice in frustrated $XY$ approximation, has a lower
bound which determined by coupling matrix of system. Formulation
of the problem in terms of properties of coupling matrix, maybe
can be used for more analytical results especially relation
between $XY$ models with different coupling matrix.

\end{document}